\documentclass[a4paper,twoside]{article}

\usepackage[utf8]{inputenc}
\usepackage[T1]{fontenc}
\usepackage{epsfig}
\usepackage{subfigure}
\usepackage{calc}
\usepackage{amssymb}
\usepackage{amstext}
\usepackage{amsmath}
\usepackage{amsthm}
\usepackage{multicol}
\usepackage{pslatex}
\usepackage{apalike}
\usepackage{url}
\usepackage{SCITEPRESS}     

\begin{document}

\title{Exploring Biofeedback with a Tangible Interface Designed for Relaxation}

\author{\authorname{Morgane Hamon\sup{1}, Rémy Ramadour\sup{1} and Jérémy Frey\sup{1}}
\affiliation{\sup{1}Ullo, 40 rue Chef de Baie, La Rochelle, France} 
\email{morgane@ullo.fr, remy@ullo.fr, jfrey@ullo.fr}
}


\abstract{Anxiety is a common health issue that can occur throughout one's existence. In this pilot study we explore an alternative technique to regulate it: biofeedback. The long-term objective is to offer an ecological device that could help people cope with anxiety, by exposing their inner state in a comprehensive manner. We propose a first iteration of this device, ``Inner Flower'', that uses heart rate to adapt a breathing guide to the user; and we investigate its efficiency and usability. Traditionally, such device requires user's full attention. We propose an ambient modality during which the device operates in the peripheral vision. Beside comparing ``Ambient'' and ``Focus'' conditions, we also compare the biofeedback with a sham feedback (fixed breathing guide). We found that the Focus group demonstrated higher relaxation and performance on a cognitive task (N-back). However, there was no noticeable effect of the Ambient feedback, and the biofeedback condition did not yield any significant difference when compared to the sham feedback. These results, while promising, highlight the pitfalls of any research related to biofeedback, where it is difficult to fully comprehend the underlying mechanisms of such technique.}

\onecolumn \maketitle \normalsize

\section{Introduction}

At times, to endure fear might be beneficial, when one is facing dangerous situations or unknown events. ``Fight or flight'' responses were proven important for survival. However if stress becomes chronic and is not treated, it can be a factor of sleep disorders or cardiovascular disease \cite{Kivimki857}. It can also lead to a pathological state of anxiety when the anticipation of stressing stimuli is sufficient to trigger the same symptoms as with the actual appearance of stimuli. Finding effective and lasting solutions to reduce stress is necessary to alleviate this public health problem, which impedes the lives of many. Treatments exist against stress and anxiety, but they might require a strong and timely involvement (i.e. therapies) or provoke side effects (i.e. drugs). Studying anxiety and offering alternative solutions to drugs (sport, yoga, mindfulness) is a growing body of research, and nowadays tools exist to let people autonomously reflect on their states and better act upon themselves.

\subsection{Biofeedback}

\begin{figure}[!ht]
\centering
\includegraphics[width=0.8\columnwidth]{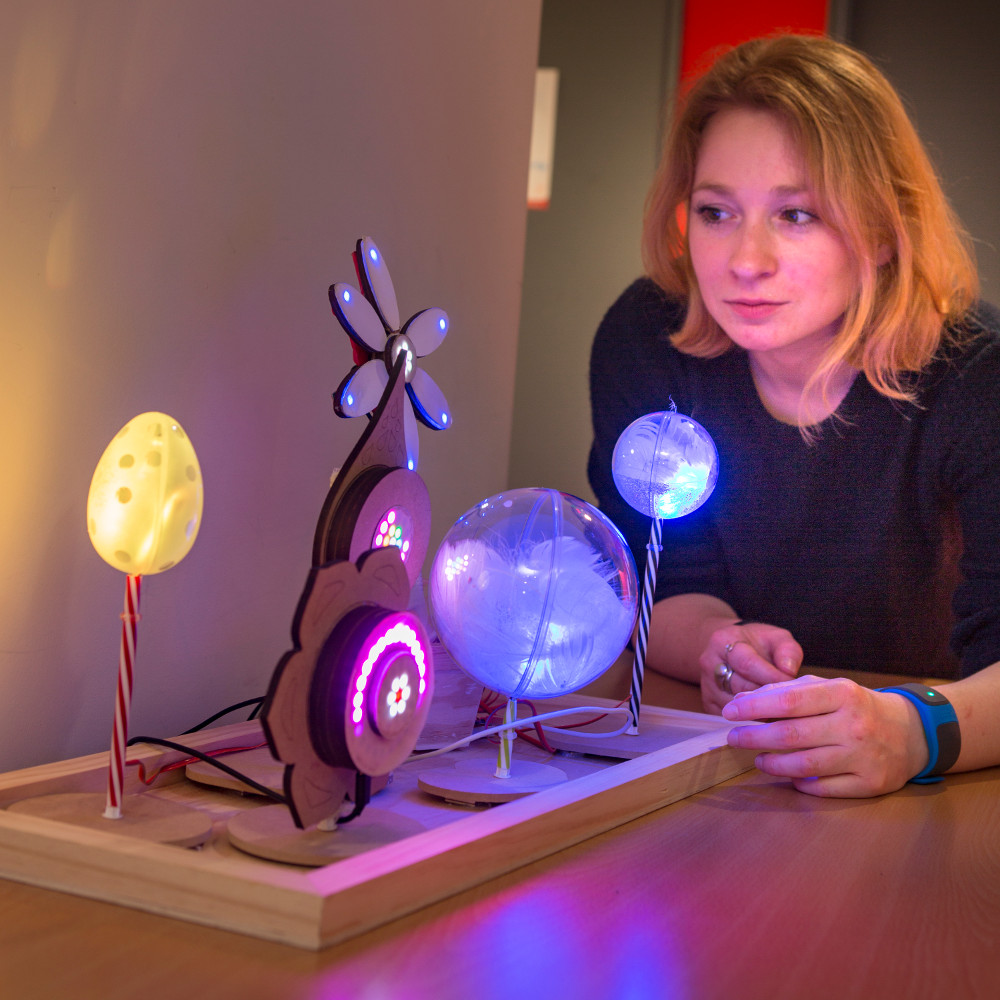}
\caption{In this example heart-rate is measured through a smartwatch; tangible interfaces serve both as breathing guides for the user and as biofeedback devices (image © Inria, photograph C. Morel).}
\label{fig:teaser}
\end{figure}

Biofeedback is a method that enables users to learn to control autonomous bodily processes. Biofeedback is part of a larger notion known as "interoception" which is defined by \cite{Farb2015} as \textit{"the sense of signals originating within the body"} or \textit{"the process of receiving, accessing and appraising internal bodily signals"}. Biofeedback relies on physiological measurements and is getting more and more popular thanks to the increasing availability of non-invasive sensors (Figure \ref{fig:teaser}). Most of the time, signals originate from respiration, electrodermal activity (EDA) or heart rate (HR) \cite{McKee2008}.

While these processes are mostly involuntary and with weak level of consciousness, through physiological sensors it becomes possible to visualize them in real-time. Users are then able to view effects of a particular behavior on their physiology . For example if a person wearing a HR sensor begins to run, HR is going to increase suddenly. Connected to a biofeedback display the increase of HR could be symbolized by a color change, a blinking light, or simply seen as a graph. This information can then be used to help users regulate their state, for example stay within a specific HR range in to prevent injuries or maximize the outcomes of the training. Biofeedback applications are diverse, from rehabilitation to stress management \cite{McKee2008}. 

It is worth noting that despite the fact that biofeedback has been investigated and applied for decades, its effectiveness has not been systematically demonstrated. Notably, even though a review such as \cite{Yucha2008} presents many studies across a variety of medical applications, there are hardly any comparisons between biofeedback and sham feedback, which is the only way to form a proper control group and pin-point the real efficacy of the technique.

Still, as a drug-free and a non-invasive approach, potential risks and side effects of Biofeedback are small. With a medical opinion it could be a compromise for people who can not or do not want to take drugs (e.g pregnant women).

\subsection{Heart Rate Variability: a Marker of Anxiety}

One of the most studied case of biofeedback for stress management is the cardiac activity. Heart Rate is under control of the autonomous nervous system (ANS) and is not constant at rest. Heart Rate Variability (HRV) -- a marker of the evolution of HR over time -- is a representative index of ANS activity. Among the variety of factors that could influence HR, a low HRV has been shown to be correlated with impaired parasympathic activity, higher anxiety, and a variety of disorders \cite{Vaschillo2006}. \cite{Tan2011} showed that veterans with Post Traumatic Stress Disorder (PTSD) had a HRV significantly lower than subjects without PTSD. The HRV was compared between veterans with PTSD who received HRV biofeedback in addition to their regular treatment and veterans without the additional biofeedback treatment. The results indicated that the group with HRV biofeedback had significantly increased their HRV while reducing symptoms of PTSD compared to the other group. Finding ways to assist people to learn to increase their own HRV and improve their health is one of the motivations of our study.

\subsection{Increasing Heart Rate Variablity}

Deep breathing is a well documented manner to increase the HRV. ``Cardiac Coherence'' is a notion according to which respiration and cardiovascular functions are synchronized. That means the HR increases during the inhalation and decreases during exhalation, and so periodically \cite{McCraty2009}. Authors proposed that 6 breaths per minute is the respiratory frequency that allows one to reach cardiac coherence and the highest amplitude in heart rate oscillations (hence the highest HRV) \cite{DeBoer1987}. Cardiac Coherence is then \textit{a priori} obtained when breathing at a 0.1Hz frequency which corresponds to six 10-second breathes per minute.

Several studies have investigated  a static breathing guidance at 6 breaths per minute to reduce stress. For example in \cite{Yu}, where authors showed that HRV was higher after the breathing exercise, while subjective impressions -- as measured by the State Trait Anxiety Index questionnaire, or STAI \cite{book} -- were not different. In \cite{Dijk2011} authors designed an immersive system which is composed of a blanket containing small vibrating motors (haptic stimulus) and headphones (audio stimulus). Haptic and audio stimuli are synchronized and generated at such a frequency that the user can follow as a breathing guidance. One of the studied frequency modality was 6 breaths/min. Some participants reported the frequency as too fast or too slow. The Authors highlighted that a poorly adapted frequency can potentially create hypo/hyperventilation. 

Even if a static guidance might, on average be optimal for the population, it is not the best way to maximize HRV for each individual. As a matter of fact,  \cite{Vaschillo2006} found that resonant frequency (equivalent to the cardiac coherence) differs according to each person. By exposing a breathing guidance from 4.5 to 6.5 breaths per minute to their participants, they managed to define the best frequency for each subject.

As such, we opted for an adapted breathing guidance to create our biofeedback; in our current study we compare it with a fixed breathing feedback (i.e. ``Dynamic'' \emph{vs} ``Static'' feedback) in order to have a better grasp on the effect of an adapted biofeedback.

\subsection{Shaping the Best Biofeedback}

\cite{Muench2008} proposed a certain type of biofeedback to achieve higher HRV and reach a resonant frequency. In this study, a graph based on HR fluctuations was created, having users inhaling until the HR reaches a ``peak'' and exhaling until it reaches a ``pit''. Because of the improved HRV as compared to breathing exercises agnostic to HR, we also designed a biofeedback device that would use HR measurements to propose an adapted breathing guide.

However, the feedback modality used to convey the information back to the user is important, and graphs are not the only way to present a breathing guidance. Actually, such kind of feedback might even impede acceptability because it could appear as being too judgmental due to its close relationship to a metric. In order to craft a more ``organic'' and yet informative biofeedback, we decided instead to rely on a physical object to present the feedback.

Indeed, thanks to recent advances in human-computer interaction, is is now possible to directly integrate digital information within users' surroundings, for example through the use of tangible interfaces. These interfaces have been proven effective to help people learn about bodily processes -- see e.g. \cite{fleck:hal-01804324}. Through tangible interfaces, biofeedback can then be more easily integrated in the natural settings of users, and become part of a specific scenario. 

\subsection{Ambient feedback}

Ambient computing contrasts with disrupting notifications. In this context ``ambient'' refers to information that is being presented in the peripheral attention of users \cite{MacLean2009}. Ambient devices do not mobilize attention. They require minimal efforts from the user and yet they provide informative feedback; rather than ``pushing'' a notification it is up to users to ``pull'' information when they require it.

Along those lines, \cite{Moraveji2011a} highlights two ways to train respiration: consciously thinking about it or learning to follow an external stimulus such as a pacing light or an auditory guide. While the former method implies focus of attention, the latter, if proven effective, could alleviate the required amount of cognitive resources \cite{Hazlewood}. 

In \cite{Moraveji2011a}, authors investigated whether a breathing guidance at the periphery of the screen during work had a influence on breath rate. As this type of guide does not require full attention they called it \textit{Peripheral Paced Respiration (PPR)}. This ambient guide can be very useful by allowing the user to fully commit to another task. The guide rate setting was 20\% below the user baseline. They showed significant difference on breath rate depending on the activation of the PPR. The same thinking led \cite{Azevedo2017} to develop a wearable device that delivers tiny vibrations on the wrist with a frequency 20\% slower than the participant resting HR. Users didn't know the function of the device and were preparing an oral presentation (to induce stress). Results showed that the control group had significantly higher anxiety according to questionnaires (STAI) and physiological data (EDA).

\cite{Schein2001} investigated in a longitudinal study if a device called BIM (for Breathe with Interactive Music) had a positive influence on Blood Pressure (BP). The BIM device creates a musical pattern which is related to the user breathing rate. A 10 minutes long quiet synthesized music recording was used as an active control. Results showed BP reduction was greater in the experimental group compared to the control, and seem to have a long-term effect (significantly different 6 months after).

Based on these various findings, we decided to investigate not only a ``Focus'' but an ``Ambient'' utilization of a tangible biofeedback device as well. In order to better frame our experimental design, we turned to previous work from psychology and physiological computing that aimed at investigating various dimensions of stress.

\subsection{Evaluating and inducing stress}

Inducing stress to evaluate short-term effects of biofeedback devices is common in the literature. To artificially put participants in a more stressed state increases the range of measurements and help to reduce variability between subjects. Different type of stress can be induced, physical (e.g. extreme heat or cold), psychological (e.g. increase in mental workload) or psychosocial (e.g. public speaking) \cite{10.3389/fnins.2014.00114}.  Psychosocial stress can be induced by faking an interview -- Trier Social Stress Test \cite{Kirschbaum1993} -- or simply asking participants to prepare a speech, as in \cite{Azevedo2017}. 

In \cite{Yu} authors induced psychological stress with a mathematical task for a ten-minutes period. Several manners exists to check if the stress inducing task had an effect. First, it can be detected by recording physiological data, for example with HRV as a marker of anxiety. EDA is also a common recorded measure -- e.g. \cite{Roseway2015} -- to detect arousal. Because physiological signals might have poor specificity, those indicators do not represent how participants are feeling, what is their state of mind. Questionnaires can compensate for this situation. To measure anxiety the most frequently used questionnaire is the STAI as in \cite{Yu}, \cite{Azevedo2017}, or \cite{10.3389/fnins.2014.00114}. In the latter work authors compared the robustness of various physiological signals to detect stress. They crossed psychosocial stress (TSS) with psychological stress, by manipulating the amount of cognitive workload endured by participants. To do so, they employed the N-back task, which leverage on memory load \cite{Owen2005}. We chose to use the N-back task as well to induce stress in our study since it is effective in doing so and since we can easily compute a performance metric to sense how participants might be affected by the exposition to a tangible biofeedback.

\subsection{Objectives}

In the present study, our first objective is to investigate if an ambient modality could have positive effects on stress. To explore this issue we compared two modalities: Ambient \emph{vs} Focus. ``Ambient'' stands for the presence of the device during the whole experiment (30min) in the peripheral visual field of the participant. ``Focus'' refers to a short (6 min) utilization in the middle of the experiment requiring full-attention.

Our second objective is to check the relevance of breathing guidance based on HR . Indeed, as we propose a new device it is important to consider that any observable effect might be due to a ``novelty effect''.  Moreover, we wanted to assess the usefulness of actually measuring physiological activity. Hence we compare a true biofeedback (``Dynamic'' condition, where the breathing guide is adapted using HR) with a sham biofeedback, or ``pseudofeedback'' (``Static'' condition, where the breathing guide is set to a fixed rate). As explained by \cite{Health1982}, a ``\textit{pseudofeedback is defined as a non contingent stimulus presented in exactly the same manner as the true biofeedback and with the intention of having subjects believe that it is true biofeedback}''.

\section{Study}

This study possesses a 2 (Attention: Ambient \emph{vs} Focus) $\times$ 2 (Biofeedback: Dynamic \emph{vs} Static) between-subject experimental plan. As a result we split our participants into four distinct groups: Ambient-Dynamic, Ambient-Static, Focus-Dynamic and Focus-Static.

Because in the Focus group the device is used only mid-experiment, the first part of the experiment in Focus serves as a control group (no device) for the investigation of attention.

Our hypotheses are:

H1. Exposition to an Ambient feedback (H1a) or to a Focus feedback (H1b) reduces psychological stress.

H2. An adapted breathing guidance with a biofeedback device (Dynamic) will reduce stress as compared to a pseudofeeback (Static).

H3. The usability resulting from the use of a biofeedback device is improved as compared to a pseudofeedback (Focus-Dynamic \emph{vs} Focus-Static).

\subsection{Inner Flower}

\begin{figure}[!ht]
\centering
\includegraphics[width=1\columnwidth]{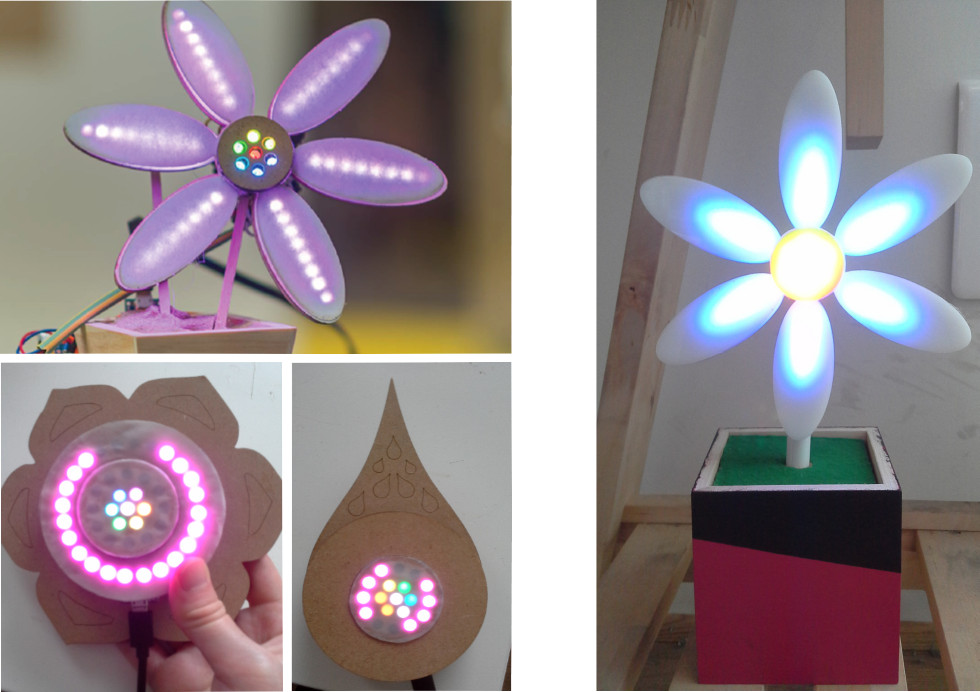}
\caption{\emph{Left}: Examples of prototypes toward an tangible biofeedback. \emph{Right}: Design of the Inner Flower used in the study}
\label{IF}
\end{figure}

In order to test our hypothesis, we iterated over various form factors to craft a tangible biofeedback that could both act as an ambient feedback and as a breathing guide. Among the biofeedback modalities that are traditionally employed -- visual, audio or haptics -- \cite{Frey2018} suggests that there is little difference regarding the effectiveness of conveying a breathing feedback. We chose to rely on visuals with color LEDs since they offered many degrees of freedom to work with (e.g. location of the lights, intensity, color, speed of the animation) and since it was easy to implement. 

We employed laser cutting, Arduino-based microcontollers (Arduino mini), and Adafruit\footnote{\url{https://www.adafruit.com/}} Neopixel LEDs to prototype several artifacts (Figure \ref{IF}). It quickly became apparent that frosted plastic was the best material to work with in order to improve the LEDs' diffusion, smooth the light and prevent an impression of a pixelated display, which would be too much reminiscence of a desktop display. Regarding the actual appearance of the object, after having explored more abstract shapes, we decided to create a somewhat stereotypical flower in order to convey a sense of well-being. Additionally, as one might want to smell the scent of a flower, it is in a way a valid proxy for breathing.

Then we had to design the feedback that would guide the breathing. While it was straightforward to pick movements as a guidance, several adjustments were required in order to smooth the animation of the LEDs that were disposed on the ``petals'' of the flower. In the final version (Figure \ref{IF}, right), the user would breathe in when the lights from of the Inner Flower move outwards and breathe out when the lights move inward. In the Static condition this animation is set to 6 breaths/min. In the Dynamic condition, the breathing rate (BR) is coupled with the heart rate of the participant. We start our investigation with a simple coupling: $BR = HR / \Delta$. The divider $\Delta$ has a default value of 15. It can be set to adapt the breathing guidance to each user during a calibration phase. During the Dynamic condition, the breathing guidance varies in real-time according to the HR of the user so as to reach cardiac coherence. For example with a HR at 90 beats per minute and a divider set to 12, the resulting breathing guide pulses 7.5 times per minute.

There is no direct representation of the HR (e.g. no light blinking at the pace of the heart rate) so as not to overwhelm users with information not related to current  usage scenario.

\subsection{Measures}

During the study we were interested in collecting two types of data: the (psychological) stress level of participants, and a usability index related to the tangible biofeedback. Stress was assessed along three dimensions: physiological activity (HR), behavioral measures (performance in a N-back task) and questionnaires (STAI). Usability was assessed through a questionnaire (USE) administrated at the end of the experiment for those groups that explicitly used the device (i.e. Focus groups).

\subsubsection{Heart rate}

Each participant was equipped with a smartwatch measuring heart rate (``Link'' from Mio\footnote{\url{https://www.mioglobal.com/}}), placed on their non-dominant hand. Mio smartwatchs employ photoplethysmography (PPG) to compute heart-rate, a technique which basically detects variations in skin's color to assess heartbeats. This solution was preferred over electrocardiography (ECG) to improve comfort and acceptance of the system, which is meant to be used in ecological settings, outside the laboratory. Even though PPG is less robust than ECG, being that the participants are steady and seated, i.e. without the risk of creating motion artifacts throughout this study, PPG is a sufficiently good sensor. During pilot studies, we validated that the instantaneous HR measured by this particular smartwatch was accurate enough to detect changes in HRV associated with deep relaxation.

Data was collected over Bluetooth, processed in real-time in the Dynamic condition and stored for further analysis. From HR measurements we focused our investigation on one index of HRV: RMSSD -- root mean sum of the squared differences. RMSSD takes as input the inter-beat interval (the inverse of the instantaneous HR measured by the smartwatch). It is one of the best indicator of cognitive workload \cite{Mehler2011}, a specific type of psychological stress induced during the experiment.

\subsubsection{N-back}

\begin{figure}[!ht]

\centering
\includegraphics[width=1\columnwidth]{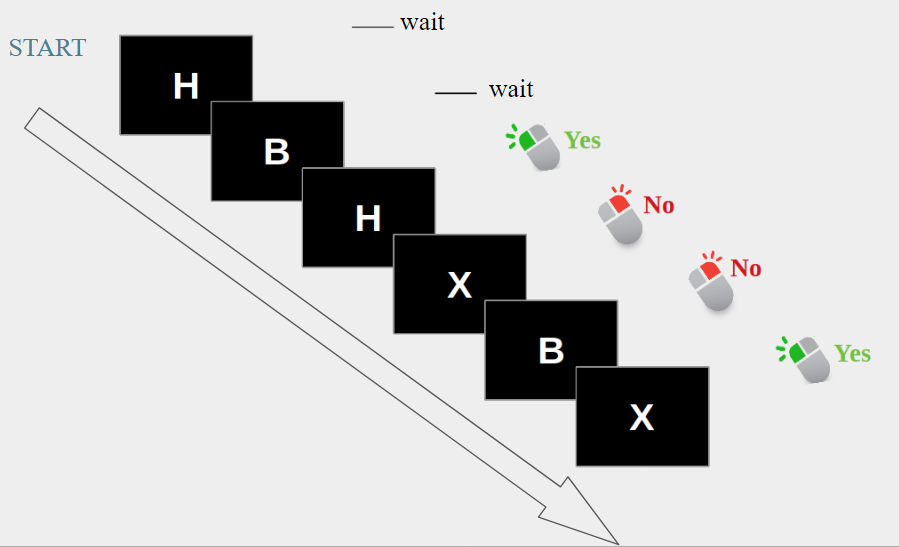}
\caption{2-back task. Participants have to click left when the displayed letter is the same as two steps before.}
\label{nback}

\end{figure}

The N-back task served both to induce psychological (cognitive) stress, and to evaluate the cognitive load of participants. The latter is revealed by calculating participants' performances. During this task, as in \cite{10.3389/fnins.2014.00114}, each letter appears on the screen for 0.5 second every 2 seconds. Participants have to determine whether the current letter is the same as the one they saw \emph{N} steps before (left click on a mouse) or not (right click). We employed a 2-back task (Figure \ref{nback}), which showed to induce a high workload level \cite{10.3389/fnins.2014.00114}. Typically, when workload increases the performance during the task decreases.

An immediate feedback is provided to inform participants whether their answers are correct or not. Two N-back tasks were presented to each participant, denoted as ``N-back1'' and ``N-back2''. Each task is comprised out of three sequences of one minute (30 letters). The first task (N-back1) includes an additional sequence at the beginning that acts as a training session. Over each N-back task we measured participants' performance (percentage of correct answers).

\subsubsection{STAI}

The State Trait Anxiety Index (STAI) enables the measurement of self-reported anxiety levels \cite{book}. The STAI Y-A version of the questionnaire measures anxiety as a state (about one's current endeavor). The classical version is composed of 20 items (e.g. ``I am tense''; ``I feel content'') that participants rate on a 4-point Likert scale. A higher score reflects a higher anxiety state. Because we administrated multiple times the STAI during the experiment, we favored a six-item short-form \cite{Marteau1992}. It was faster to fill by participants and still produced similar scores to those obtained with the full-form.

\subsubsection{Usability} 

In order to assess the usability of the device, we adapted the USE questionnaire \cite{Lund}. We removed the ``Usefulness'' subscale because it was not relevant to our study -- participants did not use the device long enough. Questions were translated to French. The resulting questionnaire comprises three subscales: ``Ease of use'', ``Ease of learning'' and ``Satisfaction''. Each is composed of three items (e.g. ``I easily remember how to use it'', ``It is simple to use''). Participants had to state their agreement to each sentence on a 4-point Likert scale (``Not at all'' ... ``Very much''). Since such usability questionnaire is poorly suited to assess an ambient device, with which users merely interact directly and/or consciously, the USE was only administrated to participants of the Focus groups.

\subsection{Participants}

\begin{figure}[!ht]
\centering
\includegraphics[width=0.9\columnwidth]{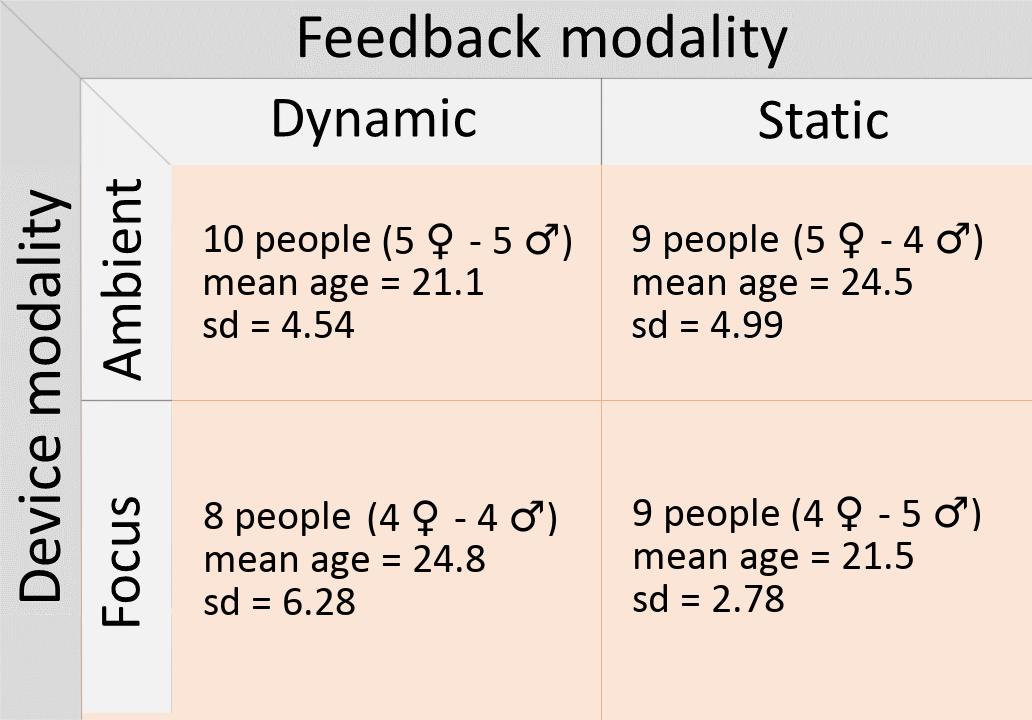}
\caption{Demographics of our groups.}
\label{participants}
\end{figure}

A total of 36 volunteers (18 females) took part in the study. Overall the mean age was 23.80 years old (SD = 4.82). The demographics per group is depicted in Figure \ref{participants}.

\subsection{Procedure}

\begin{figure}[!ht]

\centering
\includegraphics[width=1\columnwidth]{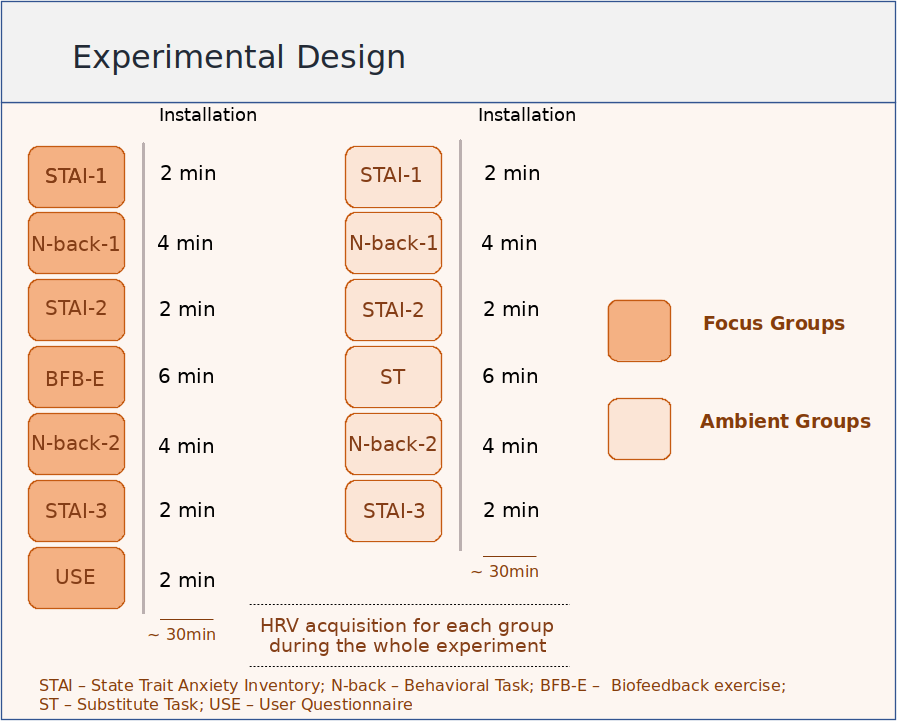}
\caption{Timeline of the experiment.}
\label{design}

\end{figure}

The timeline of the experiment is presented Figure \ref{design}. The experiment took place in a quiet room, deprived of distractions, and we detail it step by step as follows.

Upon entering the room and sitting at a table, participants signed a consent form. Afterwards we proceeded to explain briefly how the experiment would take place and equipped participants with the smartwatch. The Inner Flower was then switched on and its core functionalities were explained to participants. In the Focus-Dynamic group, a calibration phase occurred in order to determine a pace that would suit users (i.e. obtaining a breathing guide that would not appear too fast nor too slow). After the calibration, the device was switched off. On the other hand, the Focus-Static group had no calibration, hence the device was simply switched off. Lastly in the Ambient groups the device was left active on the side of the table, in the peripheral vision of participants.

Afterwards, to induce psychological stress and control for the effect of the N-back task  participants fulfilled a first STAI (STAI-1), performed a first N-back task (N-back1) and finally fulfilled a second STAI (STAI-2). At this point of the experiment, by comparing Ambient groups (active device on the side) with Focus groups (devices turned off) we are able to investigate the effect of an ambient breathing guide.

In order to assess how much the Inner Flower would affect HRV while employed as an explicit breathing guide, further in the experiment we switched-on the device in the Focus groups and participants carried out the breathing exercise. In the Ambient groups participants performed a substitute task instead; they read the short story ``The Oval Portrait'' by Edgar Allan Poe. Both tasks lasted 6 minutes and were designed to equally involve users' attention. At the end of the task, the device was switched-off in the Focus groups.

Then, all participants performed a second N-back task (N-back2). Their performance, during this test would enable us to assess the efficacy of the Inner Flower as a tool to reduce psychological stress and improve cognitive availability.

Finally, participants filled out a third STAI (STAI-3). Additionally, in the Focus groups, participants answered the USE questionnaire.

\section{Results}

Due to the nature of the data and the modest number of participants per group, we used non-parametric statistical tests to analyze the data. For studying HRV, N-back and STAI, we performed resampling (permutation) statistics using the Minque\footnote{\url{https://cran.r-project.org/web/packages/minque/minque.pdf}} package from R. Answers to the USE questionnaire were analyzed with a Wilcoxon rank sum test. When applicable, we tested for the effect of each main factor (Attention, Biofeedback and moment of measurement) as well as for the interaction of thereof.

\subsection{HRV}

\begin{figure}[!ht]

\centering
\includegraphics[width=1\columnwidth]{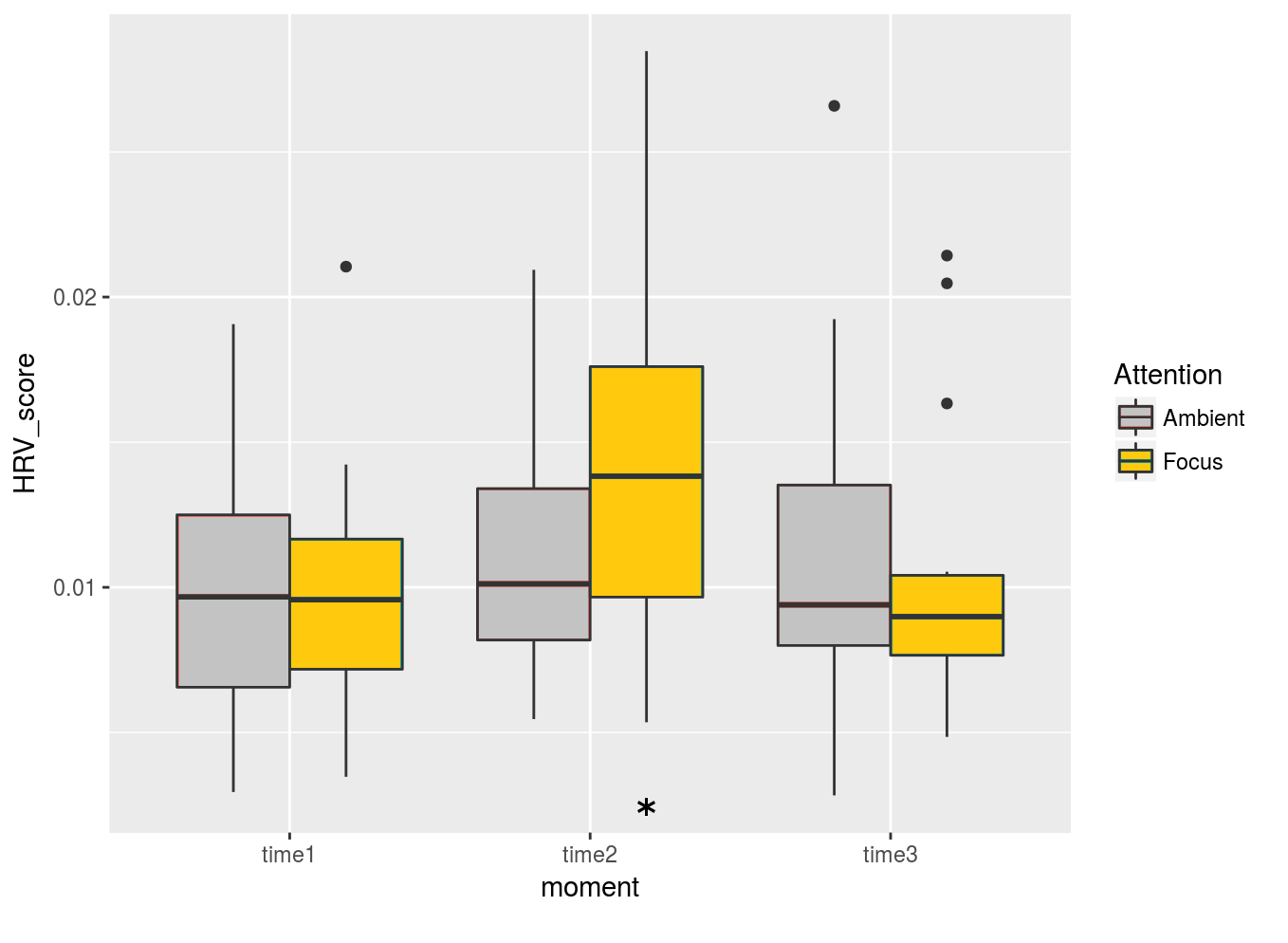}
\caption{HRV across Attention factor and time (i.e. during N-back1, exercise, N-back2). ``*'': p-value < 0.05.}
\label{hrv}

\end{figure}

HRV was computed during both N-back tasks (\emph{time1} and \emph{time3}) and during the breathing exercise (\emph{time2}).
There was a significant interaction between the Attention factor and time, with a difference between Focus during the breathing task (M=1.47$\times 10^{-2}$, SD=0.67$\times 10^{-2}$) and the rest of the conditions (M=1.07$\times 10^{-2}$, SD=0.48$\times 10^{-2}$, p < 0.05, Figure \ref{hrv}). Across other factors and interactions there were no significant differences in HRV.

Note that due to technical issues the HR data is incomplete for 5 participants (out of 36).

\subsection{N-back}

\begin{figure}[!ht]

\centering
\includegraphics[width=1\columnwidth]{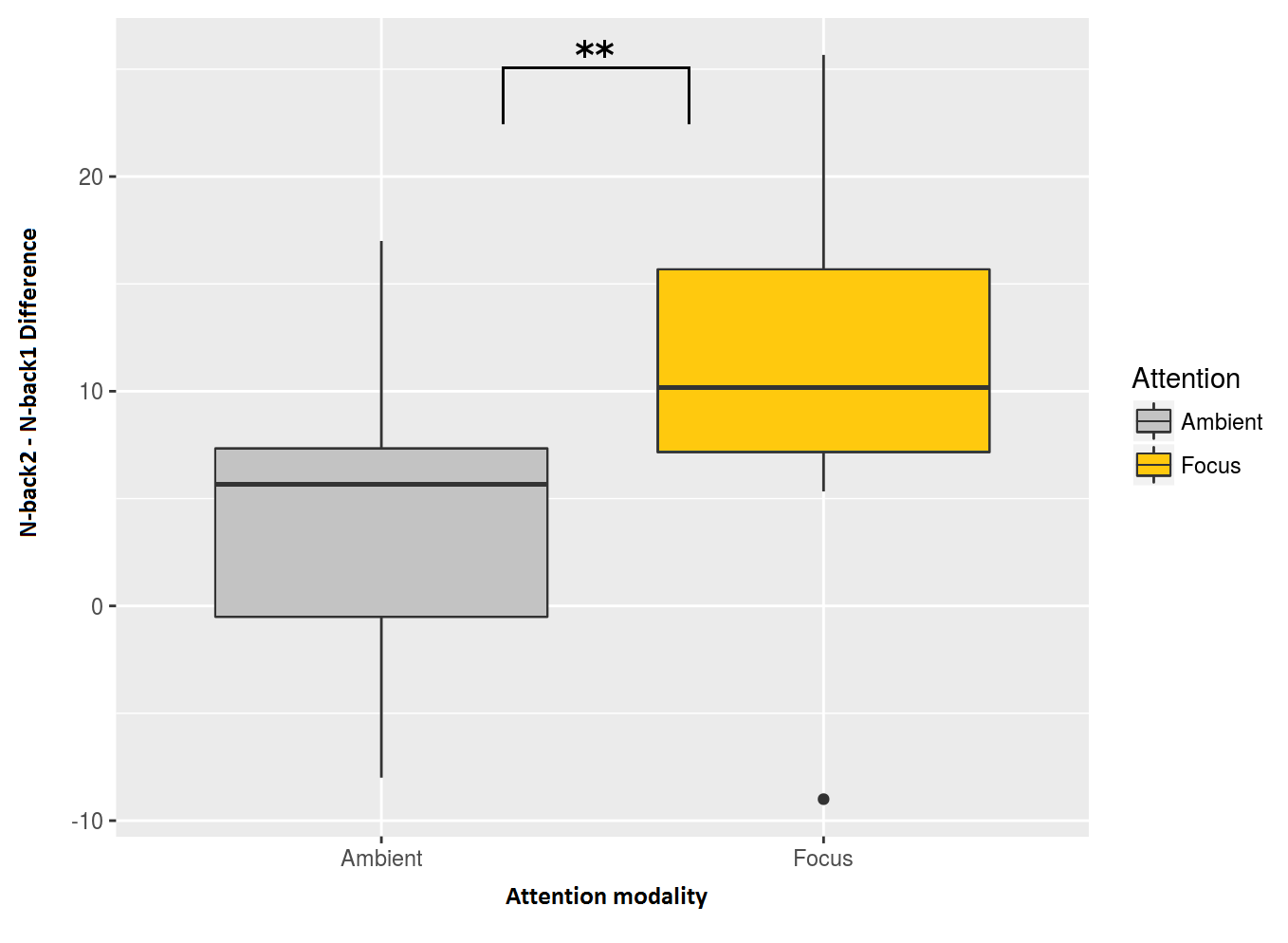}
\caption{Evolution of the performance between the first and the second N-back task. Performance increased further in the Focus groups. ``**'': p-value < 0.01.}
\label{Nback}

\end{figure}

We investigated the evolution of the percentage of accuracy during the N-back between the first and the second test (i.e. N-back2 - N-back1). Overall performance increased between those two tests; with a significant effect of the Attention factor. Performance increased more sharply in Focus groups (M=+11.23pp, SD= 8.36) as compared to Ambient groups (M=+4.21pp, SD= 7.03, p<0.01, Figure \ref{Nback}).  Across other factors and interactions there was no significant differences in the evolution of N-back performance.

\subsection{STAI}

\begin{figure}[!ht]

\centering
\includegraphics[width=1\columnwidth]{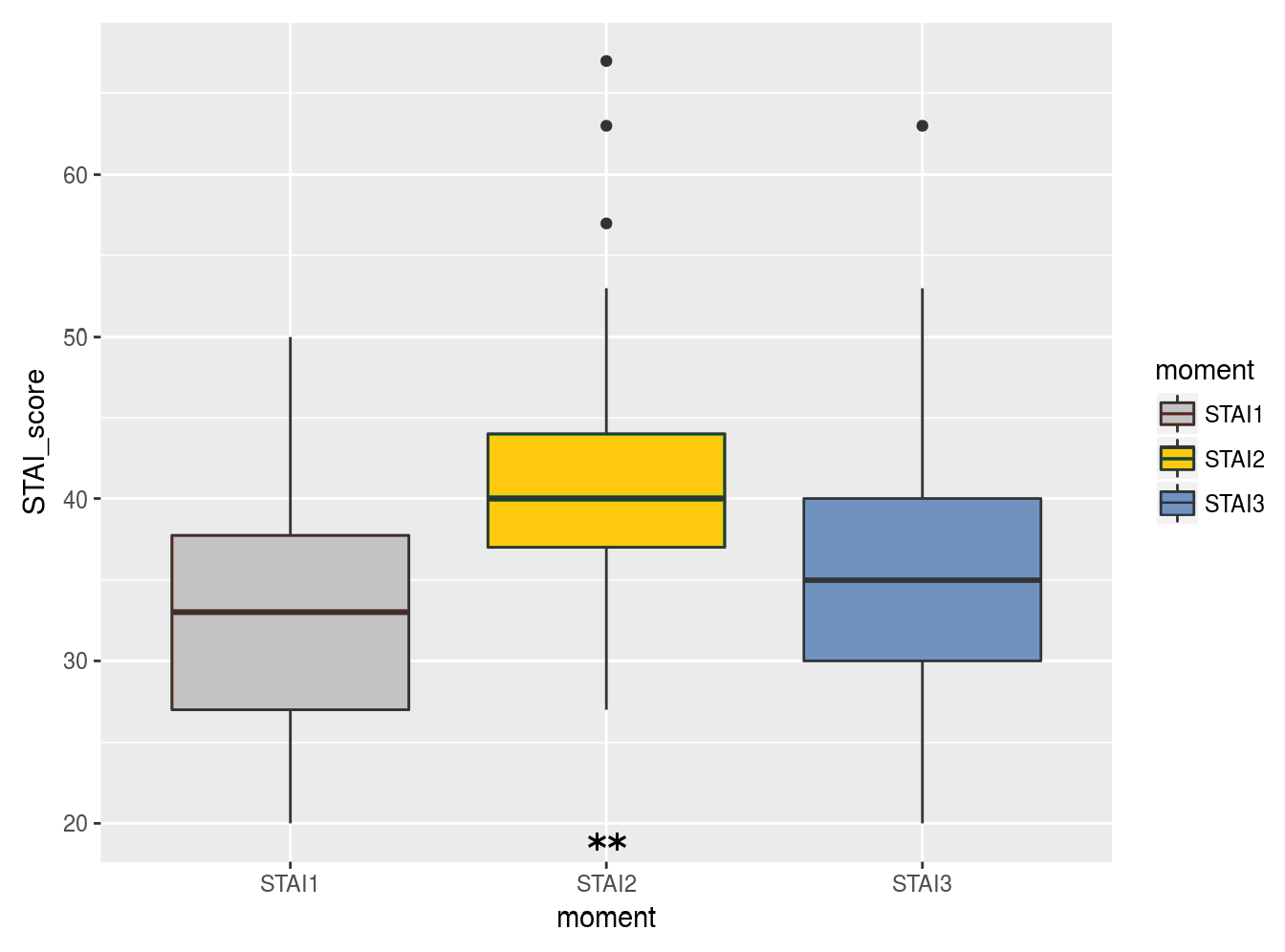}\newline
\includegraphics[width=1\columnwidth]{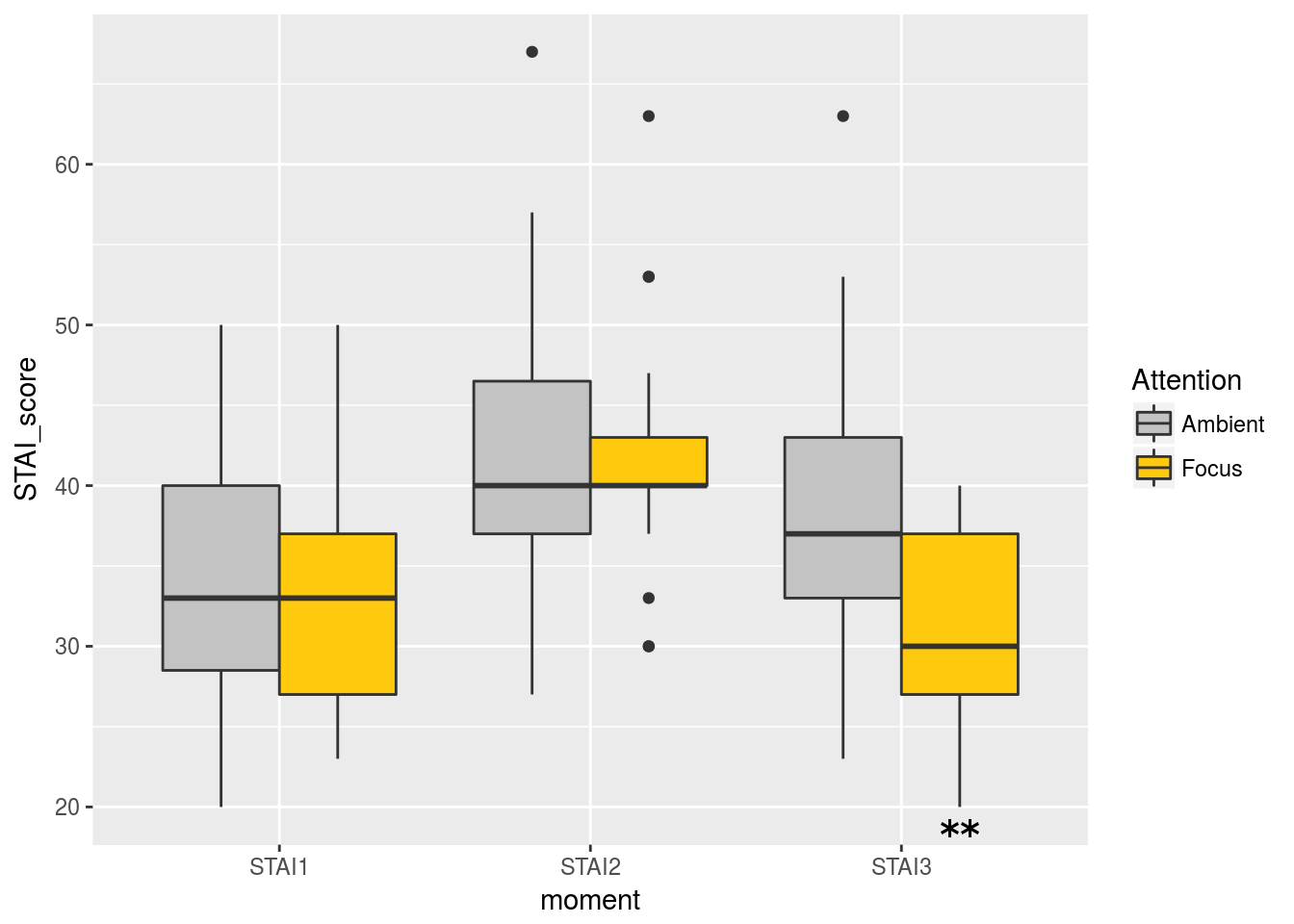}
\caption{\emph{Top}: STAI scores across time (lower score: less anxiety). \emph{Bottom}: STAI scores across Attention factor and time. ``**'': p-value < 0.01.}
\label{STAI}

\end{figure}

There was a significant effect of time. STAI-2 scores (after the first N-back task) were higher (M=42.4, SD=9.08, p < 0.01) when compared to the other scores (M=34.8, SD=8.57, Figure \ref{STAI}, top). There was a significant interaction between Attention and time, with lower scores in the Focus groups in STAI-3 at the end of the experiment (M=30.8, SD=5.92), when compared to the rest of the conditions. (M, 38.5, SD=9.48, p < 0.01, Figure \ref{STAI}, bottom). Across other factors and interactions there was no significant differences.

\subsection{USE Questionnaire}

As mentioned earlier the USE questionnaires were fulfilled only in Focus groups. There was no significant difference between Static (M=32.8, SD= 2.92) and Dynamic (M=28.9, SD=6.79, p=0.26) groups. 

\section{Discussion}

The N-back task had a significant effect on the perceived anxiety, as the STAI-2 was higher than the STAI-1. This replicates the validity of such task to induce stress. 

As opposed to previous work -- e.g. \cite{Yu}, which employed arithmetic tasks between breathing exercises -- we did not measure a noticeable effect of the Ambient condition and could not validate H1a. There was no significant difference within any marker of stress (HRV, N-back performance, STAI) between participants with an active ambient device and participants with no device (beginning of the study). We would need longer experiments, or even longitudinal studies, to better grasp the influence and the dynamic of an ambient (and subtle) biofeedback. Interestingly, when asked after the experiment during informal interviews, participants in the Ambient groups reported that they did not pay attention to the device -- an indicator that at the very least it was not disruptive. Still, it is possible that such an ambient biofeedback might reduce stress even more than a short explicit breathing exercise when users are exposed for a couple hours, or over the course of several days. 

In contrast, over the 6-minute breathing exercise participants of the Focus group demonstrated a lower level of stress on all three markers -- H1b is validated. Not only did  the breathing exercise increased HRV and helped participants to feel less anxious, but it also induced a higher performance during the N-back task. This scenario mirrors the (negative) effects that might arise when a psychological test is administrated to a sensitive population in a stressful environment (e.g. elderly in a hospital). In similar situations, the use of a device alike the Inner Flower might increase patients' cognitive availability before a test and prevent biases during a psychological evaluation.

Since there was no effect of the Feedback modality over the course of the experiment, we cannot validate H2. An actual biofeedback did not perform better (or worse) than a simple breathing exercise based on a fixed breathing rate. There are however multiple factors that could explain this outcome and that would need further investigations.

First, by trying to give more freedom to users and finely adapt the breathing guide (i.e. calibration phase in Focus-Dynamic group) we might have introduced a higher between-subject variability, as we observed more variability in the Focus-Dynamic group.


Second, the N-back task might have resulted in a ceiling effect with some participants. While overall participants tended to improve their scores between N-back1 and N-back2, those who already had a good score (i.e. >80\%, n=17) had difficulties to improve afterward. This plateau is among the confounding factors that we would need to control more finely during recruitment for future work, alongside personality traits.

Third, and maybe most importantly, our choice of biofeedback might not have been perceived as an actual manifestation of user's physiology. While we did not want to overwhelm users with too many stimuli, the absence of a dedicated HR biofeedback might have impeded their sense of agency, i.e. users did not realize that they were actually ``connected'' with the device.

This latter interpretation would be on par with the fact that there was no significant difference in the usability questionnaire. With current results we cannot validate H3, since Focus-Dynamic and Focus-Static groups rated equally high the device, between 80\% and 90\%. We would expect a higher sense of agency (i.e. in Focus-Dynamic) to be reflected on ``Ease of use'' and  ``Ease of learning'' subscales of USE.

Despite the encouraging effect on relaxation and cognitive availability of the Focus groups, when users are presented with an explicit breathing guide, the absence of results between a coupled (Dynamic) and a fixed (Static) breathing guidance highlights once again how much rigor is needed when assessing the effect of physiological measures and the resulting benefit of biofeedback.

\section{Limitations and Future Work}

As shown by \cite{Hazlewood} it is difficult to evaluate ambient technology on a one-time basis because it has to be by definition blended into the environment. In the present study, conditions were maybe not ecological enough in the sense that experiments took place in a small and impersonal room. To solve these issues we are  working with a designer to better script the overall interaction and bring the study to a home studio replica. Our next experimental design will include as well a control group without any device, for a better control of the confounding variables. While it is harder to bring the technology outside the laboratory, longitudinal studies, over several days or weeks, would inform us about the required amount of time for an ambient biofeedback to become effective.

Concerning the form factor, we conducted informal interviews at the end of the experiment. Some of the participants showed interest in multi-modal feedback, for example an audio feedback to let them continue the breathing exercise while closing their eyes. In order to enable such usage we will complement existing feedback, for example with audio waves as in \cite{Dijk2011}. 

To determine the importance of physiological measurements, we hypothesize that a social usage of the device might be a way to reinforce the association with an actual biofeedback. \cite{Roseway2015} designed an ambient device embedded in the workplace that informs those around about one's state. Thanks to that colleagues were more sensitive to emotional states of others. Similar examples inspired us to create a collaborative scenario where users would try to synchronize their heart rate through the device, with a color-based HR feedback. We envision such exercise as a way to establish trust between people. The Inner Flower could then become both a tool to manage stress and a proxy for communication, for example in situations involving care takers and care givers, or when people suffer from communicative disorders.

Over the course of the study we demonstrated how a tangible device could help reduce perceived anxiety (measured with STAI) and alleviate a physiological symptom of stress (increase in HRV). Moreover, it enabled participants to improve performance in a cognitive task. Increasing cognitive availability before a psychological test is one of the most promising applications of this technology. With a usability score between 80\% and 90\% we can state that the device was appealing to users. In the future we will further investigate the influence of both attention and type of biofeedback.

\bibliographystyle{apalike}
{\small
\bibliography{biblio}}

\vfill
\end{document}